\documentclass{emulateapj}

\newcommand{\um}{$\mu$m~}
\newcommand{\ums}{$\mu$m}

\shorttitle{Infrared Redshifts of Faint Radio Sources}
\shortauthors{Weedman et al.}

\begin{document}

\title{Redshifts from Spitzer Spectra for Optically Faint, Radio Selected Infrared Sources}

\author{D. W. Weedman\altaffilmark{1}, E. Le Floc'h\altaffilmark{2}, S. J. U. Higdon\altaffilmark{1}, J. L.
  Higdon\altaffilmark{1}, J. R. Houck\altaffilmark{1}}

\altaffiltext{1}{Astronomy Department, Cornell University, Ithaca, NY 14853; dweedman@astro.cornell.edu}
\altaffiltext{2}{Steward Observatory, University of Arizona, Tucson, AZ 85721}

\begin{abstract}
  
Spectra have been obtained with the Infrared Spectrograph on the Spitzer Space Telescope for 18 optically faint sources ($R$ $\ga$ 23.9\,mag) having 
f$_{\nu}$ (24\ums) $>$ 1.0\,mJy and having radio detections at 20 cm to a limit of 115\,$\mu$Jy.  Sources are within the Spitzer First Look Survey.  Redshifts are determined for 14 sources from strong silicate absorption features (12 sources) or strong PAH emission features (2 sources), with median redshift of 2.1.  Results confirm that optically faint sources of $\sim$1 mJy at 24\,$\mu$m are typically at redshifts z $\sim$ 2, verifying the high efficiency in selecting high redshift sources based on extreme infrared to optical flux ratio, and indicate that 24\,$\mu$m sources which also have radio counterparts are not systematically different than samples chosen only by their infrared to optical flux ratios. Using the parameter q = log$[$f$_{\nu}$(24 \ums)$/$f$_{\nu}$(20 cm)$]$, 17 of the 18 sources observed have values of 0$<$q$<$1, in the range expected for starburst-powered sources, but only a few of these show strong PAH emission as expected from starbursts, with the remainder showing absorbed or power-law spectra consistent with an AGN luminosity source. This confirms previous indications that optically faint Spitzer sources with f$_{\nu}$(24\ums) $\ga$ 1.0\,mJy are predominately AGN and represent the upper end of the luminosity function of dusty sources at z $\sim$ 2. Based on the characteristics of the sources observed so far, we predict that the nature of sources selected at 24\,$\mu$m will change for f$_{\nu}$(24\ums) $\la$ 0.5\,mJy to sources dominated primarily by starbursts.  
  
\end{abstract}



\section{Introduction}

Mid-infrared imaging surveys underway with the Spitzer Space Telescope demonstrate the presence of large numbers of sources whose luminosity must arise primarily from dust re-emission and which are significantly obscured at optical wavelengths (e.g. \citet{pap04}).  The source counts are consistent with expectations derived from efforts to account for the total infrared background and have been modeled as showing the evolution of luminous, star-forming galaxies (e.g. \citet{lag04}, \citet{chr04}).  Spectroscopic and photometric redshifts also indicate that the majority of such sources have infrared luminosity powered primarily by star formation \citep{per05}. Confirming the nature of these sources and their redshift distribution is crucial to understanding the evolution of star formation and AGN activity in the universe, especially at early epochs.  A similar situation has existed for much longer in efforts to understand the nature of optically faint radio sources (e.g. \citet{ric99}, \citet{fom02}). The observation that many sub-mJy radio sources are identified with a population of faint, blue galaxies initially indicated that most optically-faint radio sources are powered primarily by star formation \citep{har00}. This had also been indicated by the correlation between infrared and radio fluxes for starburst systems (\citet{con82}, \citet{hel85}, \citet{yun01}). However, recent studies have shown that many faint ($\sim$ 100 $\mu$Jy) radio sources do not show the infrared fluxes \citep{hig05} or submillimeter detections \citep{chp04} expected for starburst systems. It is important to determine, therefore, what fraction of the faint radio source population does have firm indications of starbursts. 

A major observational challenge is to obtain redshifts or spectral diagnostics for infrared and radio sources too faint for optical spectroscopy, sources having optical magnitudes $\ga$ 24\,mag. For sources that are sufficiently bright in the infrared (f$_{\nu}$ (24\ums) $>$ 0.75\,mJy), redshifts can be determined to z $\sim$ 2.8 using the Infrared Spectrograph on Spitzer (IRS), based on strong spectral features from silicate absorption or PAH emission.  Results for the first set of sources selected only on the basis of extreme IR/optical flux ratios indicated that such sources were typically at z $\sim$ 2 and were usually similar to the absorption spectra of local sources powered by AGN (\citet{hou05}; hereinafter H05).  However, some sources selected using colors characterising starbursts \citep{yan05} or using submillimeter detections \citep{lut05} showed PAH emission characteristic of starbursts.  Submillimeter observations of Spitzer 24\,$\mu$m sources selected because of extreme IR/optical ratios did not show the submillimeter fluxes expected for a starburst-dominated sample \citep{lut05b}; instead, most of these sources typically have stronger mid-infrared fluxes, consistent with the presence of hotter dust powered by an AGN.   It is clear, therefore, that the optically faint, infrared selected population has both AGN and starburst constituents, and further samples are needed to define their redshift distributions and relative fractions in the overall luminosity function of dusty sources.

The Spitzer First Look Survey has imaged 4.4 deg$^{2}$ at 24\,$\mu$m with the Multiband Imaging Photometer (MIPS) instrument \citep{rie04}; initial results are described by \citet{mar04} although no catalogs are as yet available.  The survey also has accompanying deep imaging with the Very Large Array (VLA) for which a source catalog is available \citep{con03} and imaging in $R$ band from the National Optical Astronomy Observatory \footnote{FLS data products are available at http://ssc.spitzer.caltech.edu/fls}. The initial comparison of 24\,$\mu$m flux densities with 20 cm flux densities (a ratio parameterized by q
= log$[$f$_{\nu}$(24 \ums)$/$f$_{\nu}$(20 cm)$]$) indicated that sources which are detected in both infrared and radio cluster around a median q value of 0.8 \citep{app04}, which is the ratio expected from the previously known radio-infrared correlations for starbursts.  This apparent agreement with expectations is misleading, because it does not include the large numbers of sources detected in the radio but not in the infrared, or vice-versa.  A comparison of Spitzer and VLA 20 cm surveys to similar detection limits, although in a much smaller sky area, indicated that only 9\% of the 24\,$\mu$m sources are detected in the radio, and only 33\% of the radio sources detected in the infrared \citep{hig05}.  Taking into account these limits, radio sources have a median q that is negative whereas infrared sources have a median q of about unity.  This result was used by \citet{hig05} to conclude that the majority of radio sources in these VLA surveys are powered by AGN, whereas the majority of infrared sources are powered by starbursts.  

The result of starburst dominance for the Spitzer 24\,$\mu$m sources based on various survey comparisons and modeling of counts is inconsistent with the conclusion of H05 based on IRS spectroscopy that most optically faint sources of $\sim$ 1mJy at 24\,$\mu$m are distant ultraluminous galaxies powered by AGN.  They observed 31 sources in the Bootes field of the NOAO Deep Wide-Field Survey
(NDWFS, \citet{jan99}) having $R$ $\ga$ 25\,mag and f$_{\nu}$ (24\ums) $>$ 0.75\,mJy.  Of the 17 sources with determinable redshifts,  16 were best fit with heavily absorbed templates and have median z of 2.2.  Because the absorbed spectra for the local templates arise from objects with AGN (the prototype being Markarian 231), and because of the radio properties and luminosities of the sources, H05 interpret the heavily absorbed sources as bring primarily powered by obscured AGN. Spectroscopy of 8 sources in the Spitzer FLS selected by \citet{yan05} using color criteria targeted to select starbursts showed that 2 of the 6 sources with measurable spectral features show strong PAH emission features.  Two sources observed by \citet{lut05} selected because of previous SCUBA detections also show PAH emission.  It is clear, therefore, that the optically faint infrared population contains a variety of sources, and  selection criteria other than simply the infrared to optical ratio need to be invoked in efforts to sort and classify this population.

To extend our samples and to begin investigation of the nature of infrared/radio sources, we undertook spectroscopic observations of sources in the FLS area selected with infrared and optical criteria similar to those in our Bootes survey but with the additional criterion of a radio detection at 20 cm.

\section{Observations and Data Analysis}

Our selection criteria for spectroscopic targets were to 
use the MIPS and VLA surveys with the $R$ band optical image to inspect all sources in the VLA 20 cm catalog of the FLS area \citep{con03} having f$_{\nu}$ (24\ums) $>$ 0.75\,mJy. To do this, we obtained the 24\,$\mu$m images from the Spitzer archive and produced an FLS catalog of 24\,$\mu$m sources, reduced with the MIPS Data Analysis Tool \citep{gor05}. Point source extraction was performed using an
empirical point spread function (PSF) constructed from the brightest
objects found in the 24\,$\mu$m image and fitted to all
the sources detected in the data. The flux of
each object is derived using the scaled fitted PSF after applying a slight
correction to account for the finite size of the modeled point spread
function.  For the present study, we are only interested in examining sources with f$_{\nu}$ (24\ums) $>$ 0.75\,mJy whereas the FLS 5 sigma limit is about 0.15\,mJy; our sample is complete, therefore, for these bright sources. Our 24\,$\mu$m catalog was compared to the VLA 20 cm catalog, and all sources having VLA detections (brighter than 115\,$\mu$Jy) were selected for comparison to the NOAO $R$ band catalog, also available in the Spitzer archive, and all of the radio/infrared sources having $R$ $>$ 23.9\,mag or being optically unidentifed were examined. (Our criteria for source association were that cataloged source coordinates agree to within 2 ''.)  Our final sample consists of the 26 sources in the FLS having f$_{\nu}$ (24\ums) $>$ 1.0\,mJy which are in the 20 cm catalog and have $R$ $\ge$ 23.9\,mag.  We have obtained spectra for 18 of these sources with the IRS; the remaining 8 sources meeting these criteria were already within the program described by \citet{yan05} and not accessible to us.

Observations and results for the sources discussed in this paper are summarized in
Table 1. Coordinates listed are the 24\,$\mu$m coordinates, which were used for the IRS targets.
Spectroscopic observations were made with the IRS Short Low module in
order 1 only (SL1) and with the Long Low module in orders 1 and 2 (LL1 and
LL2), described in \citet{hou04}\footnote{The IRS was a collaborative venture between Cornell
University and Ball Aerospace Corporation funded by NASA through the
Jet Propulsion Laboratory and the Ames Research Center}.  These give low resolution spectral
coverage from $\sim$8\,\um to $\sim$35\,\ums.  Sources were normally placed on
the slits by offsetting from nearby 2MASS stars; in a few cases with no sufficiently nearby 2MASS stars, direct pointing without offsets was used successfully.   

Sources which are observed are only a few percent of the brightness of the background flux, so the background is the dominant source of noise, and accurate background subtraction is essential before extracting a source spectrum. Background
subtraction for LL1 and LL2 modules can utilize the
background observed in the off-source order; e.g., the LL1 slit provides an observation only of
background when the source is in the LL2 slit, and this background can be
subtracted from the observation when the source is in the LL1 slit.
All images when the source was in one of the two nod positions on each
slit were coadded to obtain the image containing the source spectrum for that nod position.  The background which was subtracted from this coadded source-spectrum image included coadded background images of both nod positions when the source was in the other slit and including images from the other nod position in the same slit.  This means that the background observation subtracted from a source observation includes three times the integration time as for the source.  This improves signal to noise in the subtracted background.  We experimented with the addition of background images from observations of different sources in an attempt to produce a standard background ("supersky") for application to all sources, but the background observed from source to source was not sufficiently stable to produce a reliable supersky.  
The differenced source minus background image for each nod position was used for the spectral extraction, giving two independent extractions of the spectrum for each LL order.
These two were compared to reject any highly outlying pixels in either
spectrum, and a final mean spectrum was produced.  For SL1, there was no separate background observation with the source in the SL2 slit, so background subtraction
was done by differencing coadded images of the two nod positions in SL1.  

Data were processed with version 11.0 of the SSC pipeline, and the bcd.fits files were used for our spectral extractions.  Extraction of
source spectra was done with the SMART analysis package \citep{hig04}; location of spectra on the slit for extraction was
done by individual examination because in several cases serendipitous
sources also fell on the slit, and adopted backgrounds had to exclude
these sources.  Typical extraction widths shown in the cross-dispersion profile displayed by SMART were 3 to 3.5 pixels, to minimize noise from pixels containing little source flux.  While use of an extraction window this narrow reduces the source flux somewhat compared to the calibration observations that utilize a wider extraction, the 24\,$\mu$m MIPS fluxes are available for determining normalization of extracted spectral fluxes.  It can be seen from comparison of the spectra shown in the Figures with the MIPS fluxes in Table 1 that the extracted spectra typically agree at 24\,$\mu$m to within 10 \% of the MIPS flux. Final spectra for analysis of redshifts and features were smoothed to approximately a resolution element, applying boxcar smoothing of 0.2\,$\mu$m for SL1, 0.3\,$\mu$m for LL2, and 0.4\,$\mu$m for LL1.  These resulting spectra are shown in Figures 1, 2 and 3.  

\section{Discussion}

Even in spectra of faint sources with poor S/N, spectral features with sufficiently large equivalent widths exist in the mid-infrared spectra for redshift determination.  At one extreme, these features are the strong PAH emission features characteristic of starburst galaxies.  The other extreme shows strong absorption features, the strongest being silicate absorption, with no PAH emission.  Examples of PAH spectra for galaxies observed with the IRS are in \citet{bra04} for NGC 7714, in \citet{bra05} for the mean of starbursts, and in \citet{wee05} for NGC 3079.  The prototype absorbed AGN is Markarian 231, which shows an apparent excess above the continuum centered at 8\,$\mu$m (rest frame), caused by absorption on either side, followed by a deep trough of silicate absorption centered at 9.7\ums; the IRS spectrum is discussed in \citet{wee05}. An even more extremely absorbed source is IRAS F00183-7111, with IRS spectra in \citet{spo04}. An intermediate template with both absorption and emission is Arp 220, for which we utilize unpublished IRS spectra. This is characterized primarily by absorption, but the excess at 8\,$\mu$m is accentuated by the presence of PAH 7.7\,$\mu$m emission.  

Redshifts can be determined by seeking either the 8\,$\mu$m "excess" and following absorption, or the set of strong PAH emission features.  The strongest PAH feature is at 7.7\,$\mu$m (rest frame), so a similar redshift would be derived even if a spectrum is ambiguous as to whether the strongest feature is the 8\,$\mu$m apparent excess or true PAH emission; the interpretation of the source, however, would be very different with the two alternatives.  PAH features of about half the strength of 7.7\,$\mu$m are at 6.2\,$\mu$m and 11.2\,$\mu$m.  In order to interpret a feature as PAH, we require an indication that at least one of the other PAH features is present with the correct shape and relative flux as scaled to the 7.7\,$\mu$m feature, although the 11.2\,$\mu$m feature is often redshifted out of our observed spectral range. Redshifts are estimated in two ways: in the first, we simply seek the features described above and assign a redshift based on the peaks in the spectrum.  In the second, we fit a selection of templates that attempt a formal chi-squared fit to the full spectrum so that strengths of the features relative to the continuum are also considered. These templates are a pure PAH emission spectrum such as M82 or NGC 3079, Arp 220 (with both PAH emission and deep silicate absorption), Markarian 231, and F00183-7111.  The redshifts derived with both techniques are listed in Table 1.  Based on the differences in derived redshift depending on the template used, and on the differences in redshift measures from fitting templates and from seeking only the individual features, we estimate that the uncertainty in redshift is typically 0.1.  The final classification of an object as characterized by PAH emission (em) or silicate absorption (abs) is adopted as listed in Table 1.  Spectra with a straightforward choice of best-fitting template and spectra for which no features could be found are shown in Figure 1. Spectra having definite features but with a more uncertain assignment of template fits are in Figures 2 and 3.

Because we are attempting to use the presence of PAH emission or silicate absorption as an indicator of starburst-derived luminosity or AGN-derived luminosity, respectively, it is important to determine which characterizes the spectrum even if the redshift is not affected by this choice.  This is not always an unambiguous decision, and we present in Figures 2 and 3 examples of the 7 sources in which either interpretation (all emission and no absorption, or all absorption and no emission) can provide a possible fit to the spectral features seen. We present these examples primarily as illustrations of the cautions that must be used in interpreting these IRS spectra with poor S/N; it is important not to assign high confidence to features that may not be real. Some of this ambiguity may arise because sources actually are composite and would formally best be fit with varied combinations of PAH emission and silicate absorption spectrum.  We do not feel, however, that the S/N is adequate for this to improve our results regarding the classification of absorption dominance or emission dominance, but believe instead that the ambiguities arise primarily because of the difficulty in judging if a particular spectral peak is a real feature or is noise.  The ambiguous sources which we illustrate are numbers 1, 5, 9, 11, 14, 17, and 18; their spectra and the alternative fits are shown in Figures 2 and 3.  The choice of fit (absorption or emission) for these objects depends on deciding whether seemingly strong spectral features are real rather than accidents of noise.  We display all spectra to the longest wavelength, 35\,$\mu$m, at which some signal may be meaningfully recorded for these faint sources.  We caution, however, that any large features beyond 33\,$\mu$m should receive very little weight, because noise spikes are often found at these longer wavelengths where the detector sensitivity is falling rapidly.  As examples of the uncertainties introduced in the fits, sources 1 and 5 in Figure 2 could be well fit by PAH features if the spike at $\sim$ 33\,$\mu$m is a real feature, but if it is not real, the fit cannot be PAH.  Conversely, source 11 in Figure 2 cannot be PAH if the collection of spikes between 29\,$\mu$m and 33\,$\mu$m represents real signal. Our final decision on which fit to accept from these ambiguous cases is based on our overall judgment of which spectral features are real, but we illustrate the alternative fits so that the reader may make their own judgment and use these examples for comparison to future ambiguous spectra. There is only one case where the ambiguity in fit would result in a significant difference in redshift, because of ambiguity regarding identification of features.  This is for source 17 in Figure 3.  This is the poorest fit we have, and assigning a redshift based on the Markarian 231 template requires ignoring as noise an apparent excess of flux near 31\,$\mu$m.  If this is considered as a real feature, it is consistent with a 7.7\,$\mu$m feature at a higher redshift than any source we have yet found, and which poorly explains the continuum flux at shorter wavelengths.  We illustrate source 17 as the example of the one source for which we cannot decide if it has a measurable redshift or not.  In our final classifications summarized in Table 1, we do not include a redshift for source 17 and provide only the best fit power law for this source. The fits we finally adopt for the remaining ambiguous sources are given in Table 1; these fits are judged by whether the pure absorption or pure emission templates shown in Figures 2 and 3 provide the better overall fit to the observed spectra.  Only two of the 7 ambiguous cases in Figures 2 and 3 are judged to be better fit by PAH emission whereas the remainder are assigned to the absorption fit, except for number 17 with an indeterminate redshift.

For the total set of 18 sources, we are able to assign redshifts to 14. The median redshift of our sample is 2.1 and that of H05 is 2.2. The most notable difference in the results for the present sample compared to H05 is that we find proportionally a few more sources which are dominated by PAH emission spectra.  Two of 14 sources (numbers 3 and 9) are well fit with a pure PAH spectrum, whereas only one of the 17 sources in H05 was fit by the PAH-dominated spectrum of NGC 7714. We have 3 of 14 sources best fit by Arp 220, which contains PAH emission in the template, and there were also 3 of these in H04. This slight excess of PAH sources is the only difference between the present sample and that of H05 which might be attributable to the radio selection, presumed to favor starbursts.  The majority of sources with redshifts (12 of 14) are characterized by templates dominated by absorption (including the Arp 220 template), as were 16 of the 17 in H05. For further discussion and as a working hypothesis, we assume a simple classification into AGN-powered sources or starburst-powered sources determined only by whether the source shows strong absorption or strong PAH emission, respectively. There are 4 sources without detectable features; for these, the index $\alpha$ of the power law which best fits the observed spectrum is given in Table 1. The classification we assume for these objects is that they are also AGN characterised by hot dust but at unknown redshift; the absorption may be too weak to detect, or may be at z $\ga$ 2.8, where it would be out of the detectable range of the IRS spectra.  

It is important to note an important caveat regarding the classification of AGN sources.  This arises because of the metal-poor, compact starburst SBS 0335-052,  This has proven to have a starburst spectrum in the mid-infrared which is unique among all of the starbursts observed with the IRS.  Unlike all other galaxies observed because they were previously classified as starbursts \citep{bra05}, it has no indication of PAH features (\citet{thu99},\citet{hou04b}) and is the most extreme in this respect of all blue compact dwarf starbursts so far observed with the IRS \citep{wu05}. It also shows weak silicate absorption.  As a result, its observed mid-infrared spectral characteristics are very similar to the characteristics which we assign to AGN-powered sources. Despite its unusual spectral characteristics compared to other starbursts, \citet{hun05} and \citet{hum05} suggest that SBS 0335-052 is the best local example for the spectral shape of a primordial starburst.  If this is so, and if such objects exist at luminosities and redshifts comparable to the sources we have found, the fraction of sources which we classify as AGN-powered is smaller than we have estimated.   

Adopting the AGN classifications as described, it means that our sample has either 16 of 18 AGN, if Arp 220 absorption templates are assigned as AGN, or 13 of 18 AGN, if Arp 220 templates are considered as starbursts.  With either interpretation, the majority of sources are AGN.  The comparable numbers for the H05 sample were 30 of 31 AGN, or 27 of 31 AGN.  The median 24\,$\mu$m flux density of the H05 sample is 1.1\,mJy and of our present sample is 1.3\,mJy, so both samples are very similar in this criterion.  The current sample has median $R$ $>$ 24\,mag (because 11 of 18 do not have measurable $R$, and the faintest measured is $R$ = 24.5), whereas the H05 sample has median $R$ $>$ 25\,mag (because the optical survey limit in Bootes is fainter than in the FLS), so both samples are also similar in optical magnitude selection.  The primary difference is the use of radio flux densities to define the current sample.  Because the sample was initially defined by the infrared limit, there was no constraint on the radio flux density as long as there was a radio detection.  The radio sample reaches sufficiently faint to find sources with q as large as 1.0.  The mean q for the 18 sources is 0.56.  Only one source has a negative q; this is one of the four power law sources so is consistent with being an AGN.  The median q value is a consequence of the fact that the great majority of sources in the FLS detected both at 24\,$\mu$m  and at 20 cm have approximately this value of q, a value consistent with that from starbursts \citep{app04}. There are numerous radio sources in the FLS with more negative q, as expected from AGN, but these sources are not found in large numbers in a sample requiring detection at 24\,$\mu$m above flux densities of about one mJy. Because of the flux density limits used for our selection, our criteria lead, therefore, to a sample of objects which should be dominated by dusty, obscured starbursts according to our current understanding of the infrared and radio characteristics of such starbursts.  As a result, we expected starburst spectral classifications to dominate our sample.  

In this context, it is significant that only a minority of sources show the starburst classification determined from the presence of PAH features.  The mean q for the 13 AGN (counting the power law sources), as classified in the IRS spectra, is 0.5; the mean for the 5 starbursts (counting the Arp 220 templates) is 0.6.  Though obviously based on limited statistics, this result indicates that the q value is not a reliable diagnostic of whether a dusty source will show the PAH features expected from starbursts or the absorption expected from AGN.  This is not a surprising result if we use Markarian 231 as the prototype dusty, absorbed AGN.  At z = 2, it would have observed 24\,$\mu$m flux density of 0.6\,mJy, based on the rest-frame flux density of 1.5 Jy at 8\,$\mu$m in \citet{wee05}, so this absorbed AGN would be comparable to the flux densities of the infrared sources in our sample.  More importantly, Markarian 231 would have a q value of 0.56, based on the ratio in the rest frame of f$_{\nu}$(8 \ums) compared to the f$_{\nu}$(6 cm) of 0.41 Jy from \citet{bec91}; this value of q agrees very well with the median of the sources in our sample.

\section{Summary and Conclusions}

We have observed a new sample of 18 Spitzer 24\,$\mu$m sources with the Spitzer IRS, chosen to be optically faint ($R$ $\ga$ 24\,mag) and also to have radio detections at 20 cm. The most definitive result of this study is confirmation of the conclusions of H05 that optically-faint  Spitzer sources of f$_{\nu}$(24\ums) $\sim$ 1\,mJy are systematically at high redshift.  Of the 18 sources in our sample, 12 (67 \%) have confident redshifts z $\ge$ 1.8.  Within H05, at least 15 of 31 (48 \%) have z $\ge$ 1.8.  This is extraordinarily different from the redshift distribution expected from a flux limited 24\,$\mu$m sample as predicted using current models or observations for starburst galaxies.  For example, the redshift distribution by \citet{per05} for Spitzer sources to f$_{\nu}$(24\ums) $\sim$ 0.1\,mJy, based primarily on photometric redshifts for starburst templates, has $<$ 10\% of sources with z $>$ 1.8.  The earlier models  of \citet{dol03} predicted that for 24\,$\mu$m samples to 1.5\,mJy, less than 1\% would have z $>$ 1.8. 

The redshift distribution of our optically-faint samples is also very different from the results for optically-bright samples of similar infrared flux. \citet{bro05} use a sample similar to ours in infrared flux limit (f$_{\nu}$(24\ums) $>$ 1\,mJy) but much brighter optically ($R$ $<$ 21.7).  They have optically-determined redshifts of 255 sources in the Bootes field for which 61, or 24\%, have z $>$ 1.8, compared to the 67\% which we find.  Their selection is deliberately directed toward finding type 1 quasars by selecting objects with compact optical morphologies, and the resulting luminosity functions are similar to those derived from optically-selected quasar surveys to similar optical limits.  It is clear, therefore, that an extreme infrared to optical flux ratio is a simple but powerful criterion for preferentially selecting dusty sources at high redshift.  

More challenging than determining the redshift is the question of classifying these dusty, high redshift sources as having their mid-infrared luminosities derived primarily from starbursts or AGN. This classification is essential for utilizing counts and redshifts of mid-infrared sources to determine the correct relative contribution of these two fundamental sources of luminosity in the early universe.  Our working hypothesis for this classification, based on analogies to local ultraluminous infrared galaxies, is that sources showing strong absorption features or showing power-law continua without spectral features are classified as AGN, and those sources with conspicuous PAH features are classified as starbursts.  Using this initial hypothesis, we can examine some of the consequences of our results and make predictions regarding future results.  Considering the results in Table 1, 16 of the 18 sources show absorption features or power law spectra.  For H05, the fraction is 
30 of 31.  In either sample, this implies that over 90\% of the sources are AGN.  We can determine a lower limit to the AGN fraction by allowing that sources fit by the Arp 220 template are powered primarily by starbursts, because this template does include PAH emission even though it is dominated by absorption.  In this case, 13 of the 18 sources in Table 1 are AGN, and 27 of the 31 in H05. These diagnostics using IRS spectra indicate, therefore, that a large fraction of optically-faint sources in samples chosen at 24\,$\mu$m are powered by AGN.  Furthermore, these diagnostics indicate that at least 75\% of those faint radio sources which would be attributed to star formation based on their infrared to radio flux ratios are actually powered primarily by AGN.  These results indicate that the AGN population at high redshift is higher than currently realized, in both radio and infrared samples, primarily because of a large fraction of optically obscured AGN which were previously unknown.

There is a straightforward interpretation of these results that explains comparisons to existing models of 24\,$\mu$m source counts and makes some testable 
predictions.  The population of sources we have observed has f$_{\nu}$(24\ums) $\ga$ 1\,mJy and has extreme infrared to optical ratios such that $R$ $\ga$ 24\,mag.  If we conclude, as explained above, that this population is primarily obscured AGN, it means that the population of high redshift, obscured AGN is much larger in number than the population of starbursts at mid-infrared luminosities corresponding to this flux density limit.  It is not inconsistent, therefore, that the models of counts and redshift distributions that accomodate starbursts have not predicted this population of high luminosity, high redshift sources, because such models do not include the AGN population.  This conclusion is also consistent with observations of the only known population of luminous, high redshift, dusty starbursts, which is the SCUBA or MAMBO submillimeter population \citep{chp05}.   These submillimeter sources are known to have f$_{\nu}$(24\ums) $\la$ 0.5\,mJy (\citet{chm04},\citet{fra04}). MAMBO observations of optically obscured sources selected to have f$_{\nu}$(24\ums) $>$ 1\,mJy also indicate that these sources do not have the submillimeter fluxes expected for starburst galaxies, but have relatively larger mid-infrared fluxes \citep{lut05b}.  These various observations indicate that the luminous, dusty starburst population at z $\sim$ 2 is not sufficiently luminous in the mid-infrared to dominate the samples we selected for IRS observation. The important conclusion is that the high end of the infrared luminosity function for optically-obscured, dusty sources at z $\sim$ 2 is dominated by AGN but that the relative fraction of starbursts increases for f$_{\nu}$(24\ums) $\la$ 0.5\,mJy. This leads to a prediction about expected results from future observations of optically-faint sources selected only on the basis of f$_{\nu}$(24\ums): targets observed at fainter 24\,$\mu$m flux density levels should show a higher fraction of PAH-dominated spectra than in the samples we have presented. The one such source so far observed with f$_{\nu}$(24\ums) $\sim$ 0.5\,mJy, although chosen because of its submillimeter detection, does show PAH emission \citep{lut05}.  This conclusion also predicts that source characteristics as determined from overall spectral templates which fit a wide range of wavelengths should indicate an increasing proportion of starburst sources with decreasing f$_{\nu}$(24\ums), but populations with f$_{\nu}$(24\ums) $\ga$ 1\, mJy should be dominated by AGN templates. 

If we predict that the nature of optically faint 24\,$\mu$m sources changes for f$_{\nu}$(24\ums) $\la$ 0.5\,mJy (primarily starbursts) compared to f$_{\nu}$(24\ums) $\ga$ 1.0\,mJy (primarily AGN), this also implies consequences regarding X-ray characteristics.  In particular, comparison of X-ray and infrared detections should show a difference in the typical infrared to X-ray flux ratio at these different infrared flux limits. The mean infrared to X-ray flux ratio should be smaller for the brighter infrared sources than for the fainter infrared sources, because strong X-ray sources correspond to AGN.  While comparisons of Chandra and Spitzer surveys have been made \citep{rig04}, these have been deep surveys in small areas of the sky.  As a result, there are insufficient sources in common at the relevant flux levels to test this prediction meaningfully.  For example, there are only 10 sources with X-ray detections and f$_{\nu}$(24\ums) $>$ 1.0\,mJy in these samples as displayed by \citet{alo04}, and we do not know how many of those are optically faint or at high redshift.  Our classification has defined sources with an absorbed infrared spectrum or power-law infrared spectrum as AGN.  If this classification is correct, optically faint sources having f$_{\nu}$(24\ums) $\ga$ 1\,mJy chosen using X-ray selection as an additional criterion should always show power-law or absorbed spectra and never show PAH spectra, because the X-ray criterion is a firm AGN indicator.  So far, no IRS results are available for sources chosen from combined Spitzer and Chandra surveys, but we have programs underway to obtain these. We can be optimistic, therefore, that as spectroscopic samples of Spitzer sources accumulate, we will be able to produce quantitative luminosity functions for both dusty starbursts and dusty AGN at the crucial epoch of z $\sim$ 2.

\acknowledgments
We thank T. Herter for assistance in developing the template-fitting routines used in this paper, and we thank D. Devost, G. Sloan, K. Uchida, and P. Hall for help in improving our IRS spectral analysis.  This work is based in part on observations made with the
Spitzer Space Telescope, which is operated by the Jet Propulsion
Laboratory, California Institute of Technology under NASA contract
1407. Support for this work by the IRS GTO team at Cornell University was provided by NASA through Contract
Number 1257184 issued by JPL/Caltech.

\clearpage


\newpage

\clearpage

\begin{deluxetable}{lcccccccc} 
\tablecolumns{9}
\tabletypesize{\footnotesize}
\tablecaption{Redshifts and Source Characteristics}

\tablewidth{0pc}
\tablecaption{Properties of Sources}
\tablehead{
  \colhead{Source} & \colhead{Name\tablenotemark{a}} &
  \colhead{f$_{\nu}$ (24\ums)} & \colhead{$R$} &\colhead{q\tablenotemark{b}}
   & \colhead{time\tablenotemark{c}} & \colhead{z(f)\tablenotemark{d}}&\colhead{z(t)\tablenotemark{e}} &\colhead{$\alpha$\tablenotemark{f}}\\
  \colhead{}&  \colhead{}&\colhead{(mJy)} & \colhead{(mag)} &\colhead{} & \colhead{(sec)} &\colhead{}  &\colhead {} &\colhead {}}
\startdata
1.  & SST24 J171440.57+584104.9 &   1.30 & \nodata & 0.5  & 1200 & 1.98(abs))&2.00(231) & \nodata\\
2.  & SST24 J172210.27+584617.2 &   2.28 & 23.9   & 1.0 & 720 & 1.16(abs) &1.20(231) &\nodata\\  
3.  & SST24 J172540.26+584953.4 &   1.47 & 24.2 &  1.0 & 720 & 1.94(em) &1.80(M82) &\nodata\\    
4.  & SST24 J172249.03+585027.6 &   1.12 & 24.5    & 0.7  & 1440 &\nodata &\nodata &1.5\\
5.  & SST24 J171151.01+585104.1 &   1.05 & 24.4    & 0.5 &  1440 & 1.94(abs) & 1.92(183) &\nodata\\
6.  & SST24 J171144.26+585225.8 &   1.29 & 23.9 & 0.9 &  720 & 1.25(abs) &1.22(231) &\nodata\\
7.  & SST24 J171527.08+585802.2 &   1.04 & \nodata & -0.6  & 1440 & \nodata &\nodata &2.4\\
8.  & SST24 J171205.09+591508.6 &   1.10 & \nodata & 1.0 & 1440 & 2.31(abs) &2.40(231) &\nodata\\
9.  & SST24 J171927.28+591536.5 &   2.01 & \nodata  & 0.4  & 720 & 2.25(em) &2.20(M82) &\nodata\\
10. & SST24 J170939.20+592728.3 &   1.20 & \nodata & 0.7 & 1440 & 2.50(abs)&2.52(220) &\nodata\\
11. & SST24 J172708.53+593727.7 &   1.11 & \nodata & 0.8  & 1440 & 1.96(abs)&2.00(231) &\nodata\\
12. & SST24 J171342.79+593920.3 &   1.89 & \nodata    & 0.3    & 720 & \nodata&\nodata &1.1\\
13. & SST24 J172048.02+594320.6 &   1.54 &  \nodata & 0.3 & 1200 & 2.21(abs) &2.29(183) &\nodata\\
14. & SST24 J171343.94+595714.4 &   1.03 & 24.0    & 0.4  & 1440 & 1.70(abs) &1.81(183) &\nodata\\
15. & SST24 J171057.45+600745.2 &   2.05 & 24.5    & 0.7 & 720 & 2.38(abs) &2.40(220) &\nodata\\
16. & SST24 J172448.65+601439.1 &   1.68 & \nodata   & 0.4 & 720 & 2.22(abs) &2.40(220) &\nodata\\
17. & SST24 J172001.44+602544.9 &   1.55 & \nodata    & 0.2   & 1200 & \nodata &\nodata &0.6\\
18. & SST24 J171434.56+602828.6 &   1.51 & 24.2 & 0.1  & 1200 & 2.13(abs) &2.35(183) &\nodata\\

\enddata

\tablenotetext{a}{SST24 source name derives from discovery with the
  MIPS 24$\mu$m images in the Spitzer FLS; coordinates listed are J2000 24$\mu$m positions with typical 3\,$\sigma$ uncertainty of $\pm$ 1.2\arcsec. Source 1 is extended in the 24$\mu$m image; the total flux density is about 3 mJy compared to the unresolved flux density of 1.3 mJy listed in the Table. } 
\tablenotetext{b}{q
= log$[$f$_{\nu}$(24 \ums)$/$f$_{\nu}$(20 cm)$]$}
\tablenotetext{c}{Integration time for each order of Long Low
  spectrum; integration time in Short Low order 1 is 240 s in all
  cases.}

\tablenotetext{d}{redshift z(f) is the redshift that would be assigned by locating apparent features in the spectrum and identifying these with PAH emission features (em) at 6.2$\mu$m, 7.7$\mu$m and 11.2$\mu$m (rest frame) or the continuum excess centered at 8$\mu$m arising just shortward of the 9.7$\mu$m silicate absorption feature (abs).}

\tablenotetext{e}{redshift z(t) is the redshift determined from a formal chi-squared template fit as illustrated in Figures 1, 2 and 3; templates are identified in parenthesis as pure PAH emission (M82); combination of PAH emission and silicate absorption as seen in Arp 220 (220); pure absorption as seen in Markarian 231 (231), and very deep absorption as seen in IRAS F00183-7111 (183).  We note that source 13 has been listed by Martinez-Sansigre et al. (2005) as having a weak Ly $\alpha$ line at z = 2.02. }

\tablenotetext{f}{$\alpha$ is the power law index for the featureless continuum in sources without measurable redshift.}

\end{deluxetable}

\newpage
\clearpage

%
%


\begin{figure}
\figurenum{1}
\includegraphics[scale=0.7]{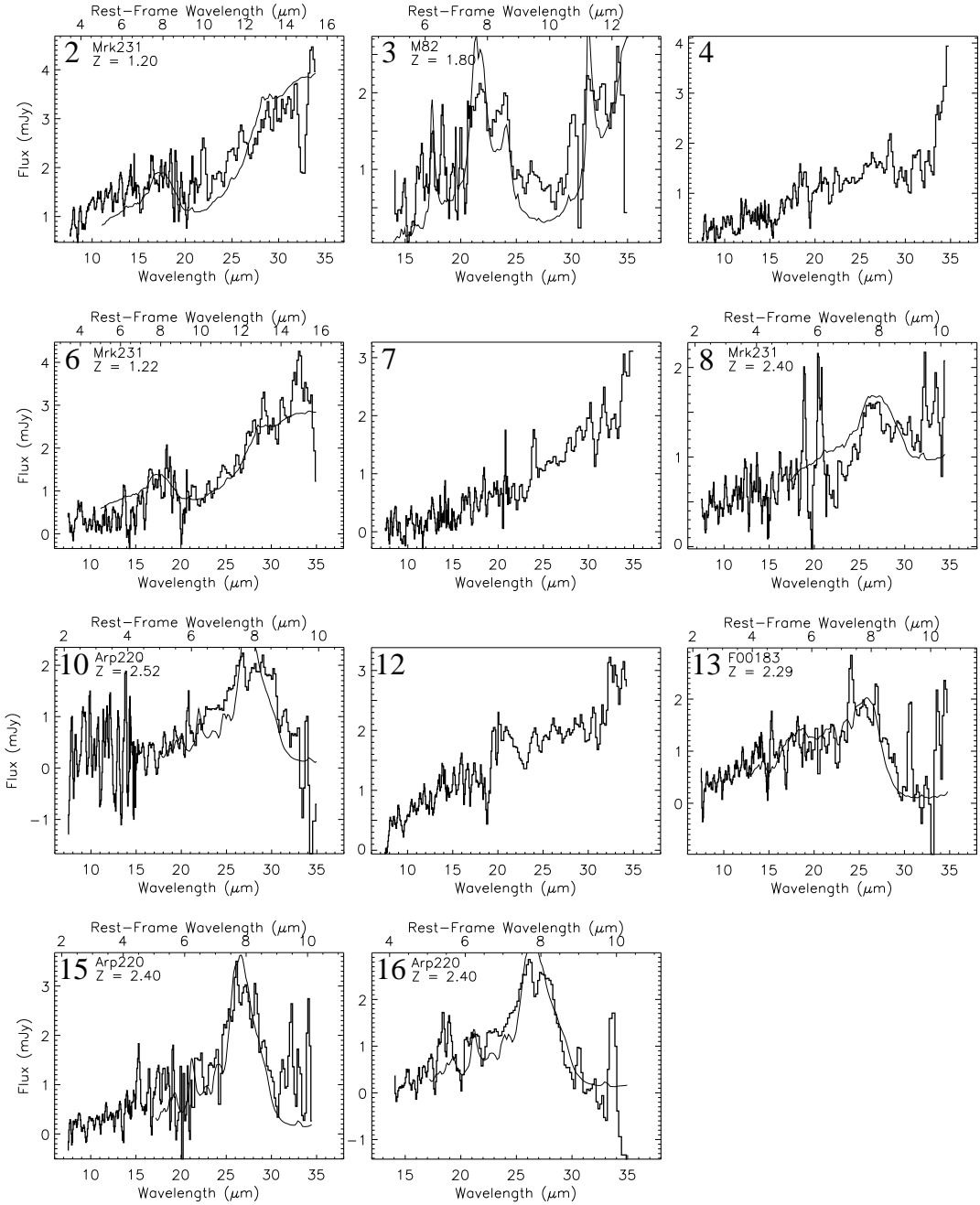}
\caption{Observed spectra smoothed to approximate resolution of individual IRS orders (histogram), best template fit (solid line), and derived redshift
  for all sources in Table 1 with unambiguous fit for redshift or with no fit; panels without a template identification or redshift are sources for which there are insufficient spectral features to derive a redshift; panels are labeled by number in Table 1.}
  
\end{figure}

\begin{figure}
  \figurenum{2}
\includegraphics[scale=0.7]{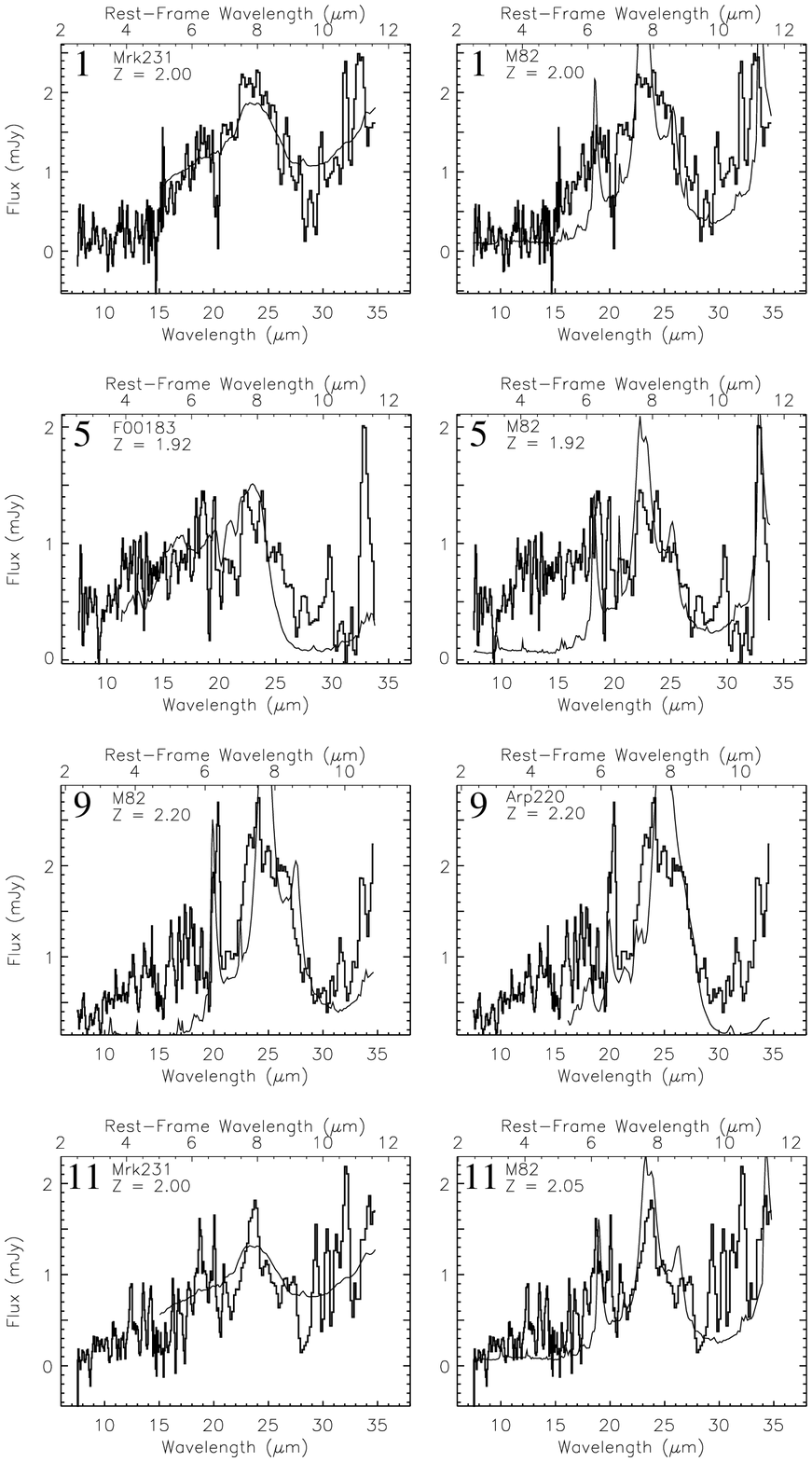}
\caption{Observed spectra smoothed to approximate resolution of individual IRS orders (histogram) and alternative template fits (solid lines) from either absorption templates or PAH emission templates and the resulting redshifts for sources for which either silicate absorption or PAH emission templates could provide a possible fit; panels are labeled by number in Table 1; first panel shown for each source is the fit selected and included in Table 1.}
\end{figure}

\begin{figure}
  \figurenum{3}
\includegraphics[scale=0.7]{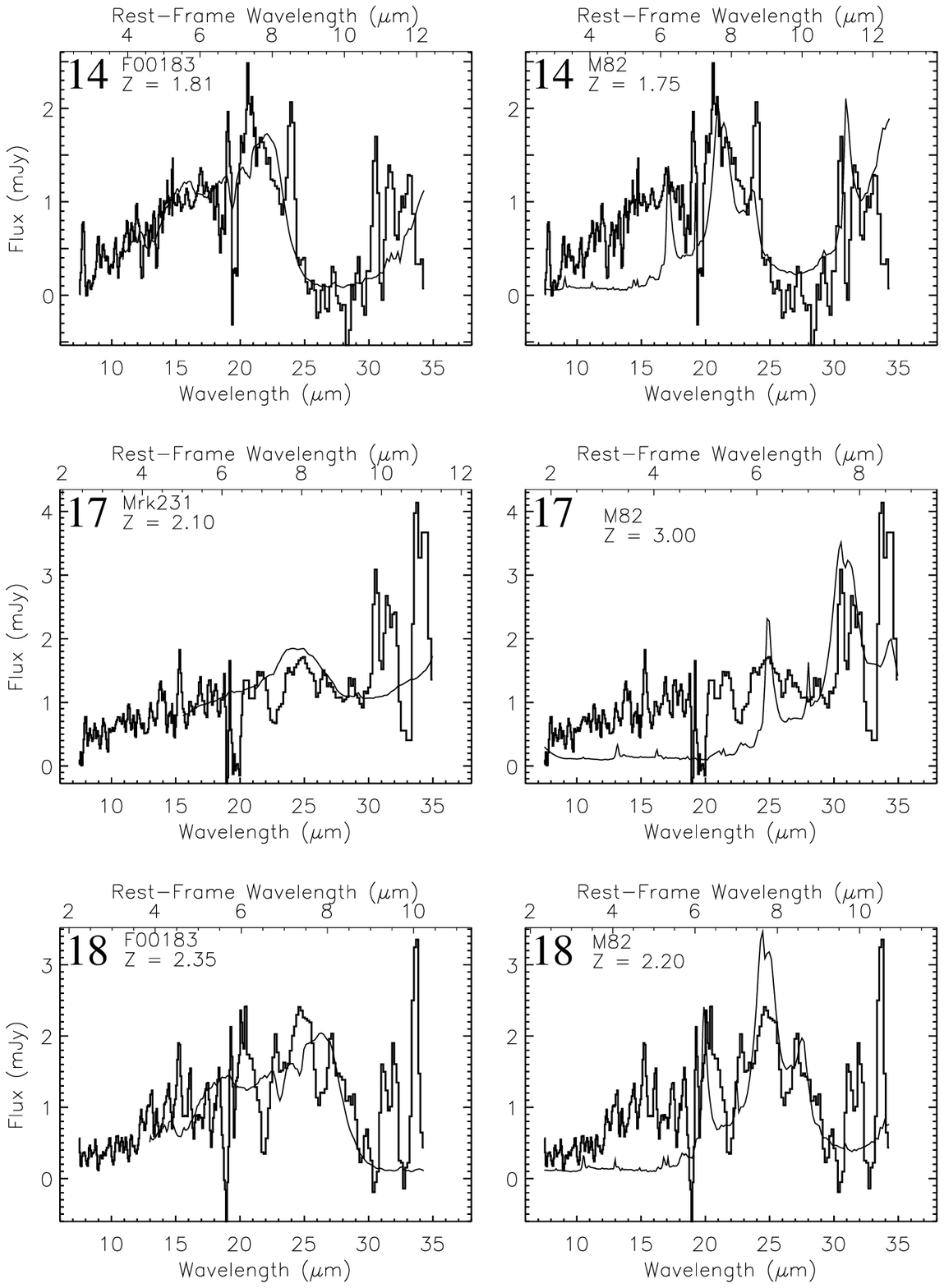}
\caption{Observed spectra smoothed to approximate resolution of individual IRS orders (histogram) and alternative template fits (solid lines) from either absorption or PAH emission and the resulting redshifts for sources for which either silicate absorption or PAH emission templates could provide a possible fit; panels are labeled by number in Table 1; first panel shown for each source is the fit selected and included in Table 1 except for source 17, for which no redshift is assigned.}
\end{figure}

\end{document}